\documentclass[onecollarge,runningheads]{svjour2}
\smartqed  
\usepackage{graphicx}
\journalname{Astrophysics and Space Science}
\begin{document}

\title{Turbulent Compressible Convection with Rotation -
    Penetration above a Convection Zone
}

\titlerunning{Penetration above rotating convection zones}        

\author{Partha S. Pal         \and
        Harinder P. Singh     \and
        Kwing L. Chan         \and
        M. P. Srivastava
}

\authorrunning{Partha S. Pal et al.} 

\institute{Partha S. Pal \at
              Department of Physics \& Astrophysics, University of Delhi,
              Delhi - 110 007, India \\
           \and
           Harinder P. Singh \at
              Department of Physics \& Astrophysics, University of Delhi,
              Delhi - 110 007, India \\
              \email{hpsingh@physics.du.ac.in}\\
            \and
            Kwing L. Chan   \at
                Department of Mathematics, Hong Kong University of Science \& Technology,
                Hong Kong, China \\
            \and
            M. P. Srivastava \at
              Department of Physics \& Astrophysics, University of Delhi,
              Delhi - 110 007, India \\
}

\date{Received: date / Accepted: date}

\maketitle

\begin{abstract}
We perform Large eddy simulations of turbulent compressible
convection in stellar-type
 convection zones by solving the
Navi\'{e}r-Stokes equations in three
   dimensions. We estimate the extent of penetration into the stable layer above a
stellar-type convection zone
   by varying the rotation rate ({\boldmath$\rm\Omega$}), the inclination of the
   rotation vector ($\theta$) and the relative stability ($S$) of the
upper stable layer
   The computational domain is a rectangular box in an f-plane
   configuration and
  is divided into two regions of unstable and stable stratification with the stable
   layer placed above the convectively unstable layer. Several
    models have been computed and the penetration distance into the stable layer above
    the convection zone is estimated by determining the position
    where time averaged kinetic energy flux has the first zero in the upper stable layer.
    The vertical grid spacing in all the model is non-uniform, and is less
    in the upper region so that the flows are better resolved in the region of interest.
      We find that the penetration distance increases as the
rotation rate increases for the case when the rotation vector is
aligned with the vertical axis. However, with the increase
   in the stability of the upper stable layer, the upward penetration distance decreases.
   Since we are not able to afford computations with finer
   resolution for all the models, we compute a number of models to
   see the effect of increased resolution on the upward penetration.
   In addition, we estimate the upper limit on the upward convective penetration from stellar convective cores.


\PACS{First \and Second \and More}
\end{abstract}

\section{Introduction}
\label{intro} Simulations of penetrative convection above a
stellar-type convection zone have been performed in two-dimensions
by Hurlburt et al. (1986), and in three-dimensions by Singh et al.
(1994, 2001) and Robinson et al. (2004). In the studies by Hurlburt
et al. and Singh et al., several models were computed to study the
overshoot of convective motions from unstable into the upper stable
layer in a non-rotating configuration. It was found that the motions
from the convective region penetrate a significant fraction of
pressure scale height into the stable layer above. Singh et al.
(1994) also found that the penetration distance above the convection
zone ($\Delta_u$) scales as $\Delta_u \sim (F_b/\rho_{ctop})^{1/3}$,
where $F_b$ is the input flux and $\rho_{ctop}$ is the density at
the top of the convection zone. Robinson et al. (2004) performed
three-dimensional simulations of the upper radiation-convection
transition layer with more realistic modeling for three subgiant
stars. For the present Sun they found the overshoot to be $0.5 H_p$,
for the $11.3$-Gyr subgiant it was $0.75 H_p$, while for a
$11.6$-Gyr subgiant it was approximately $1 H_p$, $H_p$ being the
local pressure scale height.

Simulations of the effects of rotation on convective penetration or
overshooting have been studied by Brummel et al. (2002), Ziegler \&
R\"{u}diger (2003), Browning et al. (2004), K\"{a}pyl\"{a} et al.
(2004), and Pal et al. (2007).
 Brummel et al. (2002) and Pal et al.
(2007) examined the behaviour of penetrative convection below a
convection zone under the influence of rotation by means of
three-dimensional simulations. They found that with an increase in
rotational velocity, the downward penetration decreased. A similar
behaviour was observed when the stability of the lower stable layer
was increased in a rotating configuration. Furthermore, the relative
stability parameter $S$ showed an $S^{-1/4}$ dependence on the
penetration distance implying the existence of a thermal adjustment
region in the lower stable layer rather than a nearly adiabatic
penetration zone. Ziegler \& R\"{u}diger (2003) and K\"{a}pyl\"{a}
et al. (2004), using their 3D MHD codes also found that, as a
general feature, the overshooting at the bottom decreases as a
function of increasing rotation at a given latitude.

Browning et al. (2004) performed three-dimensional simulations of
core convection within A-type stars at a range of rotation rates.
They found that as convective motions enter the stable
stratification of the radiative envelope, increasing the rotation
rate enhances the overshooting.

In this paper, we perform large eddy simulations of turbulent
convection in a {\it{f}}-plane configuration to study the effect of
rotation on the penetration distance above the convection zone. A
total of 17  two-layer (lower unstable -- upper stable) models have
been set up to examine the effect of rotation rate
({\boldmath$\rm\Omega$}) and colatitude ($\theta$) on the
penetration distance. In some models, we systematically vary the
stability of the upper stable layer to study its effect on the
penetration distance. In one model, we increase the horizontal
resolution to see if the penetration distance is altered.

In the next section, we give the essential ingredients of the
simulations and the parameters of the computed models. In Section 3,
we provide the results and their discussion. Important conclusions
of the study are listed in Section 4.

\section{Parameters of Computed Models}

The general behaviour of convective transport in a stellar-type
setting has been studied using Large Eddy Simulation (LES) approach
by several groups (Chan \& Sofia 1986, Hossain \& Mullan 1991, 1993,
Muthsam et al. 1995, Singh \& Chan 1993, Singh et al. 1994, 1995,
1996, 1998a,b, 2001 Saikia et al. 2000, Chan 2001, Pal et al. 2007).
LES, being less demanding on speed and memory as it can have a
coarser grid, allows large scale flows to be  modeled explicitly
while the smaller scales are modeled by some sort of sub-grid scale
formulation (Smagorinsky 1963).

Pal et al. (2007) numerically solved the Navi\'{e}r-Stokes equations
and incorporated the rotational effects by considering an
{\it{f}}-plane configuration. A plane parallel layer of perfect gas
was considered in a rectangular box which can be viewed as a small
portion of a spherical shell. We use a configuration that is similar
to that of Pal et al. (2007) except that the computational domain
has two layers (lower unstable -- upper stable) rather than their
three layer sandwich (stable-unstable-stable) configuration. The
choice is justified as we are interested in studying the behaviour
of penetrative convection above a convection zone rather than below
as was the aim of Pal et al. (2007). The spherical shell rotates
around the polar axis from west to east  and the angular velocity
vector {\boldmath$\rm\Omega$} points toward the north pole. We use a
right handed cartesian coordinate system in which $X$ and $Y$ denote
the horizontal direction and $Z$ denotes the upward vertical
direction. The gravity vector is denoted by {\boldmath$\rm g$} and
is along the negative $Z$ direction.  The angle between
{\boldmath$\rm\Omega$} and the $Z$ axis is denoted by $\theta$ and
the rotation vector {\boldmath$\rm\Omega$} lies in the $XZ$ plane.
The tilted rotation vector is kept constant implying a uniform
angular velocity of the sphere.

The rectangular computational domain has an aspect ratio of 1.5 and
a mesh of $35 \times 35 \times 96$ points. The top and bottom
boundaries are kept impenetrable and stress-free while the side
boundaries are made periodic. The domain is divided into two layers
with the convectively unstable layer positioned below the
convectively stable one. A constant flux {\boldmath$\rm{F}_b$} is
fed from the bottom and all the thermodynamic variables are
expressed in units which set the total depth and initial density,
pressure and temperature at the top to unity.

We use the Large Eddy Simulations (LES) approach to solve the
Navi\'{e}r-Stokes equation for an ideal gas having the ratio of
specific heats ($\gamma$) as $5/3$. The equations of the problem are
given in Pal et al. (2007). We use the Smagorinsky coefficient of
viscosity to represent the sub-grid-scale eddy viscosity in the
form:

\begin{equation}
\mu=\rho(c_{\mu}\Delta)^2(2\mathbf\sigma:\sigma)^{1/2},
\end{equation}
where $c_{\mu}$ is the Deardorff coefficient (Deardorff 1971),
$\sigma$ is the strain rate tensor and the colon sign inside the
bracket denotes tensor contraction and
$\rm\Delta^2=(\rm\Delta_x\rm\Delta_y)^{1/2}\rm\Delta_z$. The SGS
turbulent diffusivity is computed from this viscosity by assuming a
constant Prandtl number $\rm{Pr}=1/3$

\begin{table}
\caption{Physical Parameters for Models R1 to R17}\label{table:1}
\centering
\begin{tabular}{cccccrrrcc}
\hline\hline &     &       & \multicolumn{2}{c}{Polytropic Indices}&
&
        & &\multicolumn{2}{c}{Layer Thickness(PSH)} \\
\cline{4-5} \cline{9-10}\\
Models & Grid & $\rm{F_b}$ & Top & Bottom & $P_b$ & $T_b$ &$\rho_b$ & Top & Bottom \\
\hline
R1-R13 & $35 \times 35 \times 96 $ & 0.125 & 2.0 & 1.5 & 4655 & 19.5 & 238 & 6.04 & 2.40 \\
R14 & $35 \times 35 \times 96$ & 0.125 & 3.0 & 1.5 & 19767 & 17.8 & 1106 & 7.12 & 2.77 \\
R15 & $35 \times 35 \times 96$ & 0.125 & 4.0 & 1.5 & 65092 & 16.8 & 3862 & 8.01 & 3.07 \\
R16 & $35 \times 35 \times 96$ & 0.125 & 5.0 & 1.5 & 178290 & 16.1 & 11024 & 8.76 & 3.33 \\
R17 & $46 \times 46 \times 96$ & 0.125 & 2.0 & 1.5 & 4655 & 19.5 & 238 & 6.04 & 2.40 \\
\hline
\end{tabular}
\end{table}

The governing fluid equations are solved using an implicit scheme
(Chan \& Wolf 1982) and the time step is characterized by the CFL
number

\begin{equation}
N_{CFL}=C_s\Delta{t}/\Delta_{min},
\end{equation}
where $\rm\Delta_{min}$ is the minimum grid size in any direction
and $C_s={\gamma}^{1/2}$ is the dimensionless sound speed at the
top. The boundary conditions imposed are :
\begin{eqnarray}
{\mathbf {F_b}} = constant = 0.125 \,\,\,(at \,\,the\,\,
bottom),\nonumber
\end{eqnarray}
\begin{eqnarray}
T=T_t=constant \,\,\,(at \,\,the \,\,top),\nonumber
\end{eqnarray}
\begin{eqnarray}
v_z=0\,\,;\,\,\,\frac{\partial{v_x}}{\partial{z}}=\frac{\partial{v_y}}{\partial{z}}=0.\nonumber
\end{eqnarray}

\begin{figure}
\centering{\includegraphics[width=8cm]{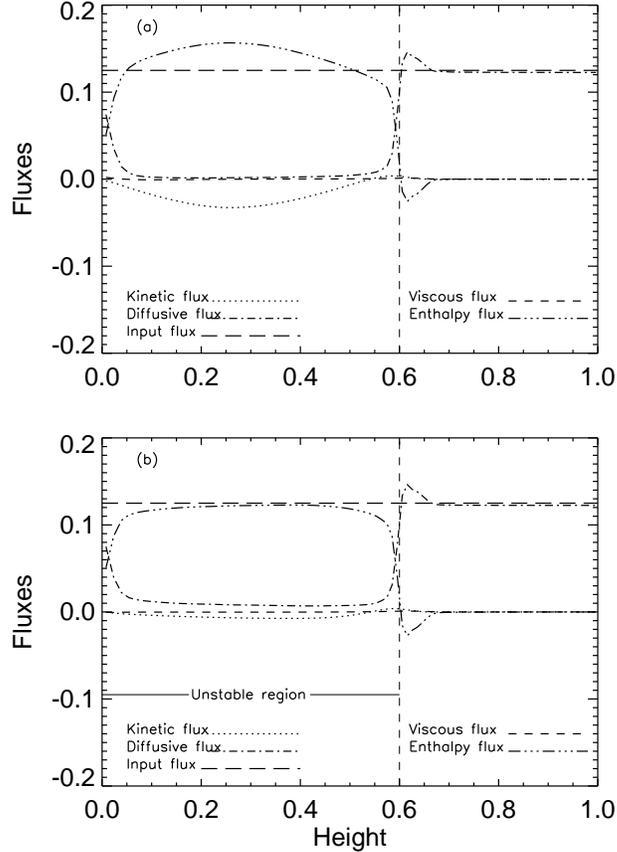}} \caption
{Profiles of the time- and horizontally-averaged energy fluxes for
(a) case R1 $(\mathbf{\Omega}=0,\, \theta=0\,^\circ)$, (b) case R7
($\mathbf{\Omega}=0.5,\, \theta=45\,^\circ$). \label{fig.1}}
\end{figure}

We have computed $17$ models and the physical parameters for the
models are listed in Table 1. The numerical parameters of the
simulations are given in Table 2. Columns (6) - (9) in Table 3 list
the values of Coriolis number ($Co$), Rossby number ($Ro$),
effective Reynolds number ($Re$), and Taylor number ($Ta$) for the
simulated cases. Following Chan (2001), they are defined as
\begin{eqnarray}
Co={\mathbf\Omega} d /\langle v^{\prime\prime} \rangle, \nonumber
\end{eqnarray}
\begin{eqnarray}
Ro=1/Co,                     \nonumber
\end{eqnarray}
\begin{eqnarray}
Re=\langle v^{\prime\prime} \rangle d/ \langle
\bar{\mu}/{\bar{\rho}} \rangle,      \nonumber
\end{eqnarray}
\begin{eqnarray}
Ta=(2{\mathbf\Omega}d^2/ \langle \bar{\mu}/\bar{\rho} \rangle)^2,
\nonumber
\end{eqnarray}
with $v^{\prime\prime} \equiv ({{v_x}^{\prime\prime}}^2+
{{v_y}^{\prime\prime}}^2+{{v_z}^{\prime\prime}}^2)^{1/2}$ where
$v_x^{\prime\prime}$ denotes the root mean square (rms) fluctuation
of $v_x$ etc. The extent of domain is denoted by $d$.

Models R1 to R13 are computed to examine the effect of rate of
rotation ({\boldmath$\rm\Omega$}) and the angle ($\theta$) between
the rotation vector and the vertical axis on the extent of
penetration below the convectively stable layer. The polytropic
index of the upper stable layer is taken to be 2.0 in all these
models (R1-R13). In models R2 to R5, $\theta$ is varied from
$22.5\,^\circ$ to $90\,^\circ$ while {\boldmath$\rm\Omega$} is fixed
at 0.25. In the second set of models R0, R10, R11 and R13,
{\boldmath$\rm\Omega$} is varied from 0 to 1 in steps of 0.25,
respectively, keeping $\theta$ fixed at $0\,^\circ$. In models R3,
R7, and R12, {\boldmath$\rm\Omega$} is 0.25, 0.5 and 1.0,
respectively, while $\theta$ is fixed at $45\,^\circ$. In the last
set of models R14 to R16, polytropic index of the upper stable layer
is varied from 3 to 5 while all the other parameters are the same as
model R7. Another model R17, having a grid of $46 \times 46 \times
96$ points and all other parameters same as R7
({\boldmath$\rm\Omega$} $= 0.5,\, \theta = 45\,^\circ$) is computed
to examine the effect of increase in the number of grid points
(resolution) in the horizontal plane.

As described , four models, namely, R7, R14, R15 and R16 have been
set-up to examine the effect of stability of the upper stable layer
on the penetration height of these rotating configurations. Each of
these four models have different polytropic indices in the upper
stable layer, namely, 2, 3, 4 and 5 for the cases R7, R14, R15 and
R16, respectively. For all these four models, the values of
{\boldmath$\rm\Omega$} and $\theta$ are kept constant at 0.5 and
$45\,^\circ$ respectively. Similar to the relative stability
parameter $S$ for the lower stable layer (Hurlburt et al. 1994,
Singh et al. 1995, Pal et al. 2007), we can define a relative
stability parameter for the upper stable layer for these four cases
as:

\begin{equation}
S=\frac{m_{i}-m_a}{m_{2}-m_a}\,\,;\,\, m_i=2,3,4,5,
\end{equation}
where $m_{i}$ denote the polytropic indices of the upper stable
layer for various cases, $m_{2}$ is the polytropic index of the
upper stable layer for our case R7 and $m_a=1/(\gamma-1)$ is the
adiabatic index. Here, we have taken $\gamma=c_p/c_v$ as 5/3. Thus,
for our reference case R7 the relative stability parameter $S$ is
equal to unity and for cases R14, R15 and R16 it comes out to be 3,
5 and 7, respectively.

\begin{table}
\caption{Numerical Parameters for the 17 models}              
\label{table:2}
\centering                                      
\begin{tabular}{ccccccc}
\hline\hline                        
Models & $g$ & $C_{\mu}$ & $\rm\Delta t$ & $t$ & $N_{CFL}$ & $\rm\Delta_{min}$ \\
\hline
R1-R15,R17 &  50 & 0.2 & 0.00054795 & 1698 &  0.1 & 0.007074 \\
R16        & 50  & 0.2 & 0.00027397 & 1698 & 0.05 & 0.007074 \\
\hline                 
\end{tabular}
\end{table}

\begin{table}
\caption{Dynamical Parameters of the Computed Models}            
\label{table:3}
\centering                                      
\begin{tabular}{ccccccccccc}
\hline\hline                        
Model&{\boldmath$\rm\Omega$}  & $\theta/\pi$ & $\theta$&
$\langle{v}^{\prime\prime}\rangle$
& $Co$ & $Ro$ & $Re$ & $Ta$ & $\rm\Delta_u$ & $\rm\Delta_p$ (PSH)\\
\hline
R1 & 0    & 0   & $0^o$    & 0.074 & $0.000$& $\infty$ & 1088.883 &  0.00E+0 & 0.055& 0.384 \\
R2 & 0.25 & 1/8 & $22.5^o$ & 0.059 & 4.231  & 0.236 & 894.394 &  5.728E+7 &0.057& 0.399  \\
R3 & 0.25 & 1/4 & $45^o$   & 0.064 & 3.914  & 0.255 & 947.021 &  5.502E+7 &0.057& 0.399  \\
R4 & 0.25 & 3/8 & $67.5^o$ & 0.070 & 3.579  & 0.279 & 1022.810 &  5.360E+7 &0.051& 0.354 \\
R5 & 0.25 & 1/2 & $90^o$   & 0.066 & 3.776  & 0.265 & 957.403 &  5.227E+7 &0.031& 0.210  \\
R6 & 0.50 & 1/4 & $22.5^o$ & 0.056 & 8.980  & 0.111 & 840.540 &  2.279E+8 &0.060& 0.422 \\
R7 & 0.50 & 1/4 & $45^o$   & 0.059 & 8.432  & 0.119 & 879.430 &  2.200E+8 &0.055& 0.385 \\
R8 & 0.50 & 1/4 & $67.5^o$ & 0.067 & 7.411  & 0.135 & 975.684 &  2.091E+8 &0.052& 0.362 \\
R9 & 0.50 & 1/2 & $90^o$   & 0.064 & 7.795  & 0.128 & 875.589 &  1.864E+8 &0.022& 0.147 \\
R10& 0.25 & 0   & $0^o$    & 0.058 & 4.278  & 0.234 & 889.006 &  5.787E+7 &0.058& 0.407 \\
R11& 0.50 & 0   & $0^o$    & 0.054 & 9.283  & 0.108 & 822.673 &  2.331E+8 & 0.063& 0.445 \\
R12& 1.0  & 1/4 & $45^o$   & 0.069 & 14.474 & 0.069 & 1011.623 &  8.576E+8 &0.052& 0.363 \\
R13& 1.0  & 0   & $0^o$    & 0.050 & 20.004 & 0.050 & 768.311 &  9.448E+8 &0.063& 0.446 \\
R14& 0.50 & 1/4 & $45^o$   & 0.036 & 13.714 & 0.073 & 876.052 &  5.774E+8 & 0.013& 0.111 \\
R15& 0.50 & 1/4 & $45^o$   & 0.026 & 19.013 & 0.053 & 895.473 &  1.160E+9 &0.006& 0.061 \\
R16& 0.50 & 1/4 & $45^o$   & 0.019 & 25.469 & 0.039 & 799.152 &   1.659E+8 &0.005& 0.058\\
R17& 0.50 & 1/4 & $45^o$   & 0.062 & 7.971  & 0.126 & 1107.676 & 3.118E+8 & 0.049& 0.340 \\
\hline                 
\end{tabular}
\end{table}

\begin{figure}
\centering{\includegraphics[width=8cm]{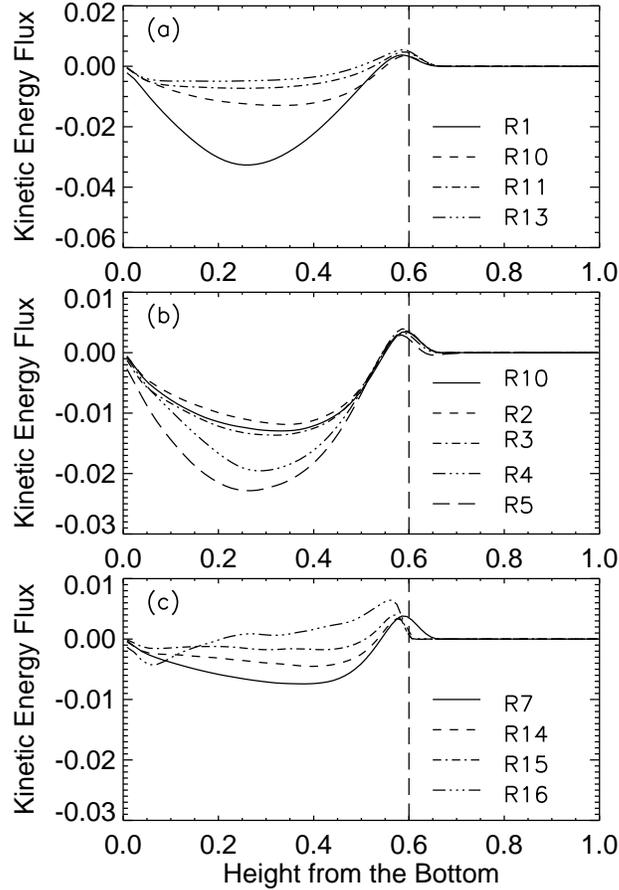}}
\caption{Distribution of time- and horizontally averaged kinetic
energy fluxes with height for models (a) R1, R10, R11 and R13. The
corresponding values of angular rotational velocity
({\boldmath$\rm\Omega$}) are 0, 0.25, 0.50 and 1.0, respectively.
For these four cases the angle ($\theta$) between the rotation
vector and the vertical axis is $0\,^{\circ}$, (b) R10, R2, R3, R4
and R5. All these models have {\boldmath$\rm\Omega$} fixed at 0.25
while $\theta$ changes from $0\,^{\circ}$ to $90\,^{\circ}$ in steps
of $22.5\,^{\circ}$, (c) R7, R14, R15 and R16. The polytropic
indices in the upper stable layer for these four models are,
respectively, 2, 3, 4 and 5. The corresponding relative stability
parameters are 1, 3, 5 and 7, respectively. All the four models have
{$\mathbf{\Omega}=0.5$}  and $\theta=45\,^{\circ}$. \label{fig.2}}
\end{figure}

\begin{figure}
\centering{\includegraphics[width=10.9cm]{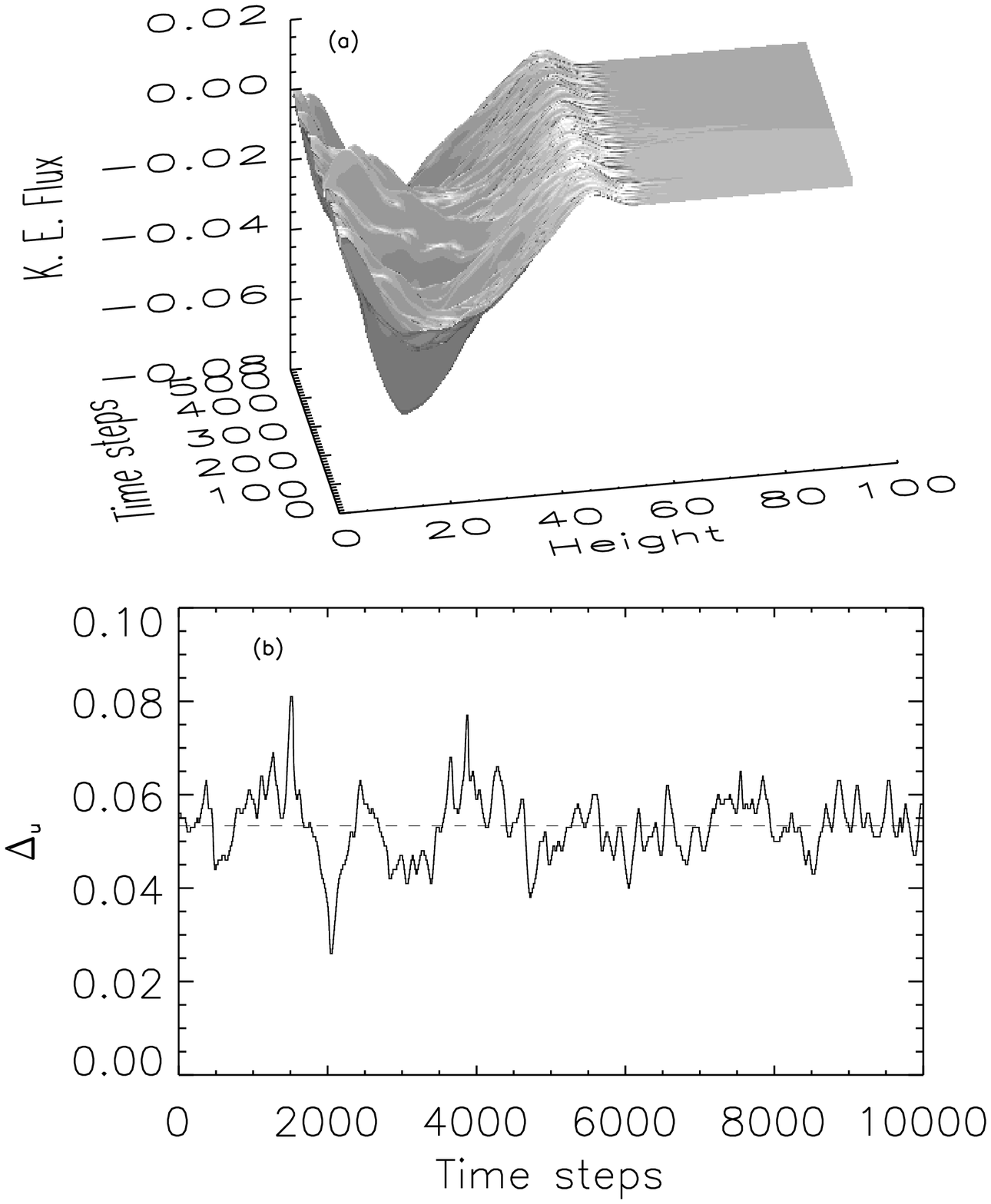}} \caption{(a)
Horizontally averaged kinetic energy flux as a function of time.
Total of 1,00,000 profiles were collected out of which 500 (every
200th) are plotted for case R1.  The height is denoted in the grid
points starting from one at the bottom (b) Penetration height
($\rm\Delta_u$) with time for model R1 for 1,00,000 time steps
(computed at every 10th time step). \label{fig3}}
\end{figure}

\section{Results and Discussion}
All 17 models computed in this study have a two layer configuration
in which the stable region is placed above the convectively unstable
region. Model R1 is non-rotating while models R2 to R17 are all
rotating, differentiated by a set of values of rate of rotation
({\boldmath$\rm\Omega$}), the angle between the $Z$-axis and the
rotation vector ($\theta$) and the polytropic index (or stability)
of the upper stable layer. Tables 1 and 2 list the physical
parameters of all the models.

The interface of the unstable-stable layer is located at a height of
0.6 from the bottom. This implies that the thickness of the lower
stable layer is 0.6 or 60\% of the total domain of computation while
the upper stable layer has a thickness of 0.4 corresponding to 40\%
of the total domain. The total domain of computation contains about
8.44 pressure scale heights (PSH) for models R1-R13 and R17 with the
upper stable layer containing 6.04 PSH while the lower convective
layer contains 2.40 PSH. For models R14, R15 and R16 the total
domain has around 9.89, 11.08 and 12.09 PSH, respectively with the
upper stable layer containing around 7.12, 8.01 and 8.76 PSH,
respectively.

It may be noticed that although the upper stable layer occupies only
40\% of the domain, it contains more than 70\% of the PSH. This is
because the length of the PSH decreases as we go from bottom to the
top. The density contrast ($\rho_{bottom}/\rho_{top}$) of models
R1-R13 is 238.0 while the temperature contrast
($T_{bottom}/T_{top}$) is 19.5 (cf. Table 1). For model R16 the
density contrast is the maximum (11024).

After the fluid has thermally relaxed, it is further evolved for
another $1,00,000$ time steps and the time and horizontal averages
are taken for quantities of interest. Figure 1 shows the
distribution of various energy fluxes with height for a number of
models. The calculation of the extent of the penetration into the
upper stable layer has been based on time- and horizontally averaged
kinetic flux ($\rm{F_k}$). Such a choice is obvious as the kinetic
flux is directly related with the motions and the profile is also
convenient for estimation of the extent of penetration (Hurlburt et
al. 1986, Hurlburt et al. 1994, Singh et al. 1994,1995, Saikia et
al. 2000, Pal et al. 2007). We illustrate this point by plotting the
distribution of kinetic energy flux for several sets of models in
Fig. 2. The kinetic energy flux is negative in most of the
convectively unstable region. Near the interface in the unstable
region it increases and becomes positive and then falls to zero in
the upper stable region. Ideally, the extent of penetration should
correspond to the first zero of the kinetic energy flux in the
stable layer above the unstable-stable interface. However, we define
the penetration height ($\rm\Delta_u$) to be the distance from the
interface of the unstable-upper stable layer to where $\rm{F_k}$ has
fallen to a value of $0.0001$ in the upper stable layer. The extent
of penetration distance ($\rm\Delta_u$) and penetration distance in
PSH ($\rm\Delta_p$) for all the 17 models are given in Table 3.

In Fig. 3(a), we show the horizontally averaged kinetic energy flux
as a function of depth and time for model R1.  We have plotted $500$
profiles of $\rm{F_k}$ which  corresponds to every 200th time step.
In Fig. 3(b), we show a time series of penetration distance
$\rm\Delta_u$ calculated by using the criterion outlined above. The
mean penetration is  indicated by the horizontal dashed line.

We now describe the effect of varying various parameters, e.g.,
{\boldmath$\rm\Omega$}$,\, \theta$, $S$, and the horizontal
resolution on the penetration distance above the convectively
unstable region.

\begin{figure}
\centering{\includegraphics[width=10.9cm]{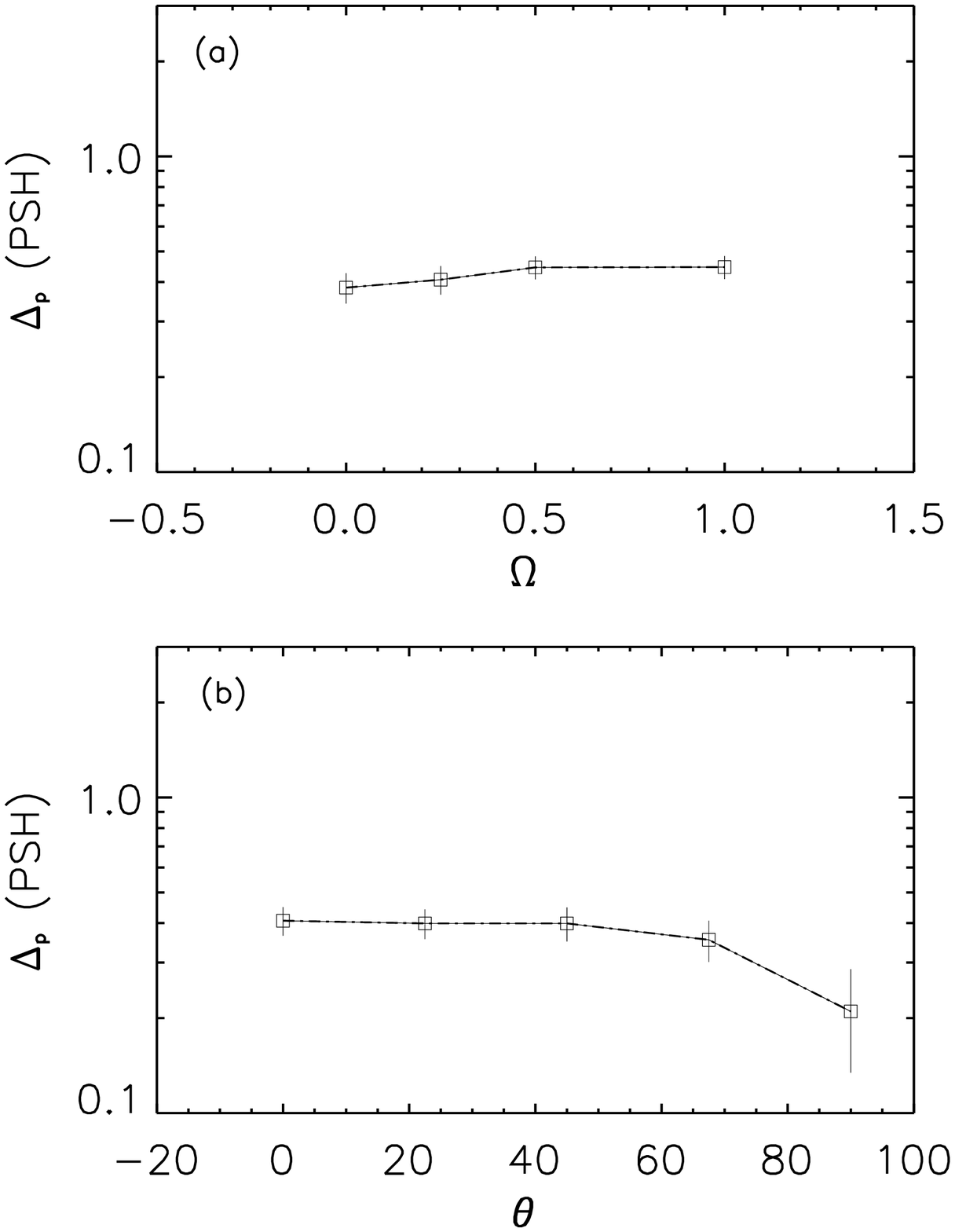}}
\caption{Dependence of penetration height in PSH ($\rm\Delta_p$)
with (a) {\boldmath$\rm\Omega$} varying from 0, 0.25, 0.50 and 1.0
for the models R1, R10, R11 and R13, (b) colatitude ($\theta$)
varying from $0\,^{\circ}$  to $90\,^{\circ}$ in steps of
$22.5\,^{\circ}$ for models R10, R2, R3, R4 and R5 having
{$\mathbf{\Omega}=0.25$}. In this as well as in all the subsequent
figures, error bars denote the standard deviation in $\rm\Delta_p$
computed for $1,00,000$ time steps after the fluid has thermally
relaxed. \label{fig.4}}
\end{figure}

\subsection{Dependence of penetration height on \boldmath$\rm\Omega$}

Two sets of models R1, R10,  R11 and R13 and R3, R7 and R12 have
been computed to examine the effect of rate of rotation on the
penetration distance. In models R1, R10,  R11 and R13, the rotation
rate ({\boldmath$\rm\Omega$}) is systematically increased from 0 to
1.0 in steps of 0.25 (cf. Table 3). In all these four models, the
rotation vector ($\theta$) is kept at $0\,^\circ$, implying that the
rotation vector coincides with the $Z$-axis or the vertical
direction. Figure 4(a) shows the dependence of penetration distance
on {\boldmath$\rm\Omega$} for these four models. In model R1, which
is a non-rotating case ({\boldmath$\rm\Omega$}$\,=0$), the
penetration distance $\rm\Delta_u=0.055$ or 0.384 PSH. As the
rotation rate increases to 0.25 in model R10, the distance
$\rm\Delta_u$ increases to 0.058 which corresponds to 0.407 PSH. For
models R11 and R13 having {\boldmath$\rm\Omega$}$\,=0.5$ and 1.0
respectively, the penetration distance increases further to 0.445
PSH and 0.446 PSH respectively.

In the second set of models R3, R7 and R12, while the rotation rates
are, respectively, 0.25, 0.5 and 1.0, $\theta$ is changed to
$45\,^\circ$. We find that unlike in the first set of models (with
$\theta = 0\,^\circ$), the penetration distance decreases with
increase in the rotation rate. For example, the upward penetration
decreases from 0.399 PSH (Model R7, {\boldmath$\rm\Omega$} $= 0.25$)
to 0.363 PSH (Model R12, {\boldmath$\rm\Omega$} $= 1.0$). The
buoyancy-breaking is accentuated as rotation rate is increased when
$\theta$ is $45\,^\circ$.

The penetrative behaviour is markedly different from the case of
downward penetration (Pal et al. 2007), in which the penetration
distance ($\Delta_d$) decreased with increase in the rotation rate,
for any fixed value of colatitude $\theta$.

\begin{figure}
\centering{\includegraphics[width=8cm]{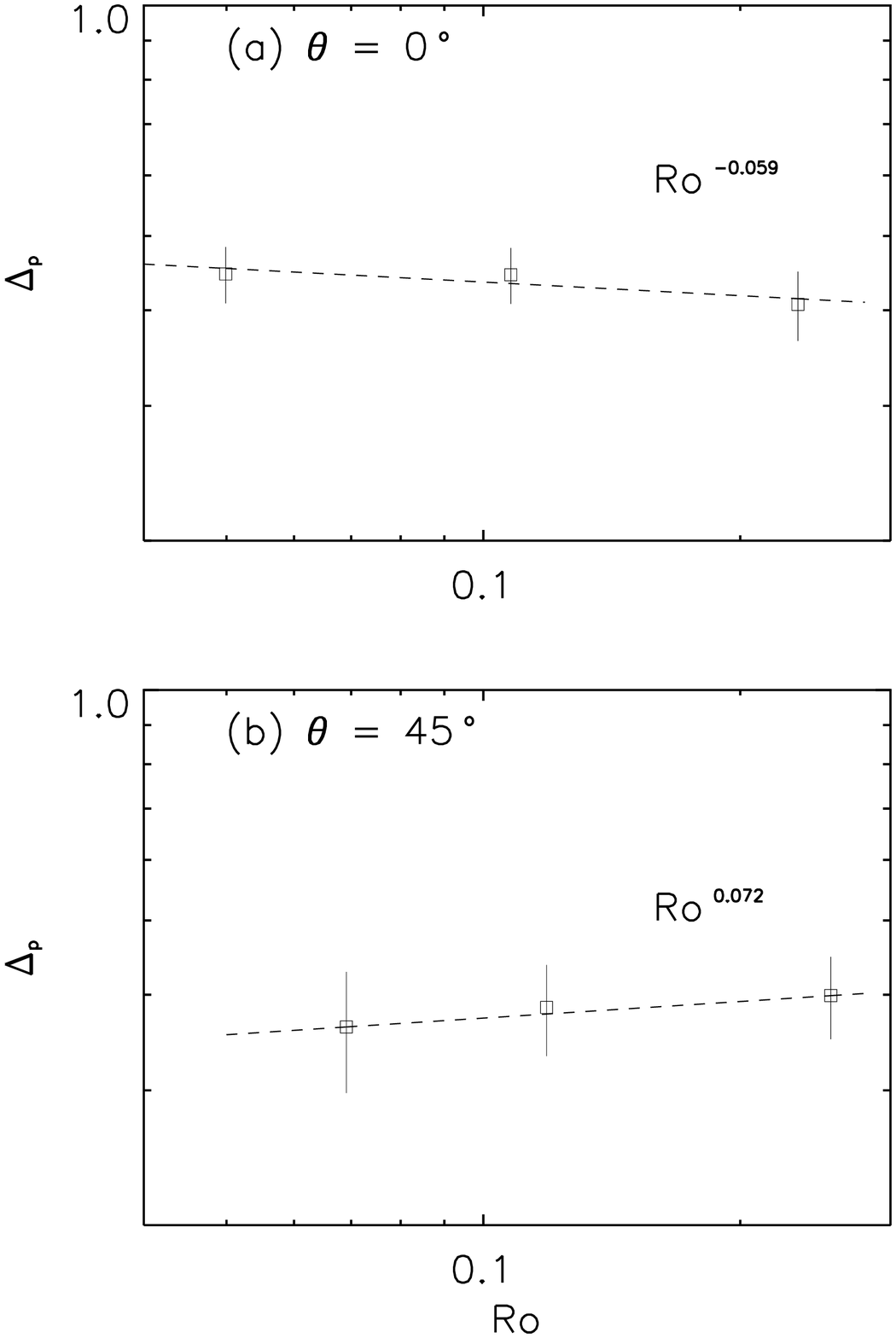}}
\caption{Variation of penetration height with Rossby number for (a)
three models R10, R11 and R13 having $\mathbf{\Omega}=0.25,\,0.50$
and $1.0$, respectively, and $\theta=0\,^\circ$, and (b) three
models R3, R7 and R12 having $\mathbf{\Omega}=0.25,\,0.50$ and
$1.0$, respectively, and $\theta=45\,^\circ$ The dashed lines
represent scaling laws: (a) $Ro^{-0.059}$ and (b) $Ro^{0.072}$.
\label{fig.5}}
\end{figure}

\begin{figure}
\centering{\includegraphics[width=8cm]{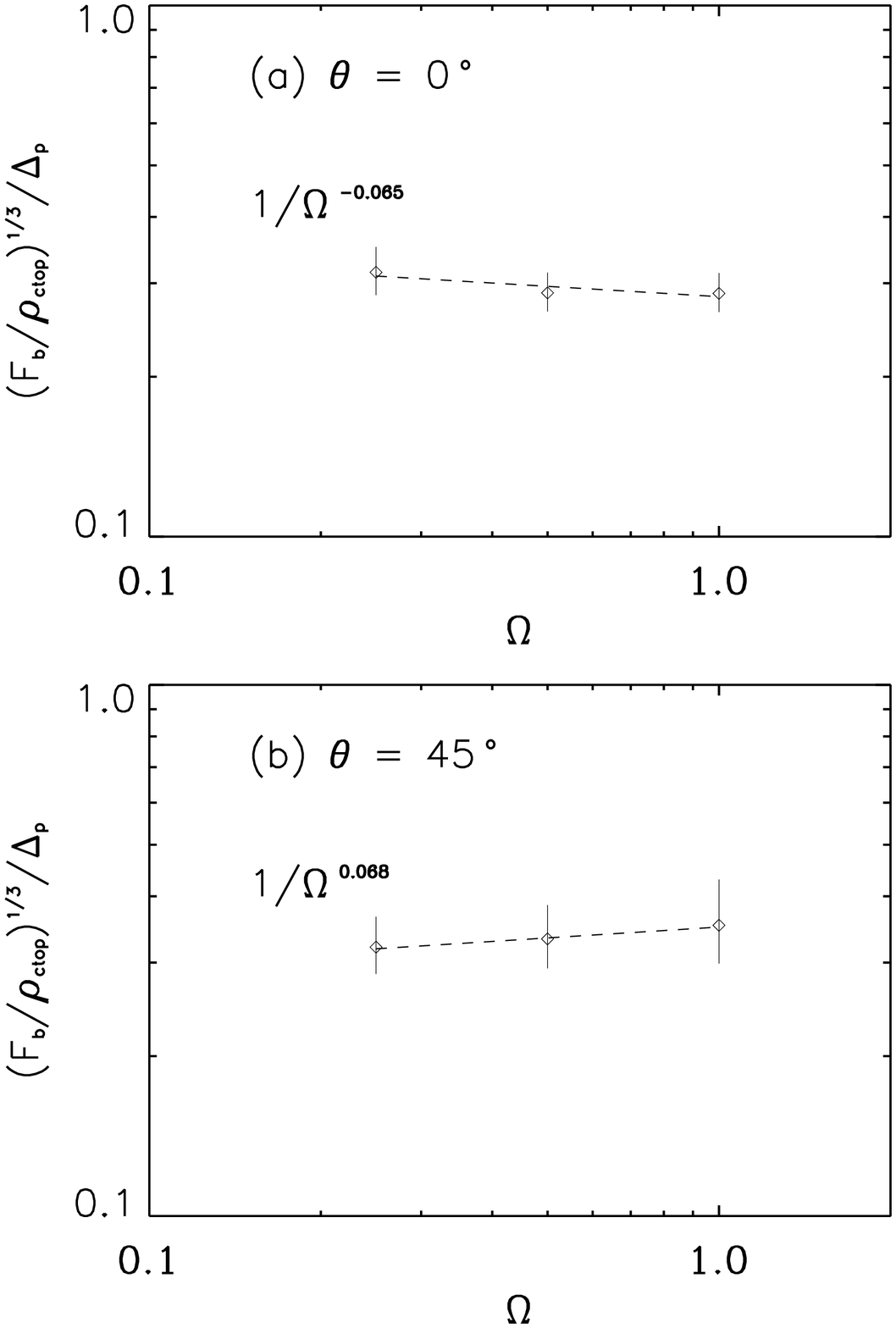}} \caption{Plot
of $(\rm{F_b}/\rho_{ctop})^{1/3}/\Delta_p$ for different $\Omega$
for (a) cases R10, R11 and R13 having $\theta=0\,^\circ$ and (b)
cases R3, R7, R12 having $\theta=45\,^\circ$. The dashed line shows
the scaling relationship (a) $1/\Omega^{-0.065}$ and (b)
$1/\Omega^{0.068}$ for the two sets of models. \label{fig.6}}
\end{figure}

For the two sets of models with $\theta=0\,^\circ$ and
$\theta=45\,^\circ$, we examine the behaviour of pentration height
with Rossby number. Figure 5(a) shows the variation of penetration
height in PSH ($\rm\Delta_p$) with Rossby number ($Ro$) for three
cases R10, R11 and R13 having $\theta=0\,^\circ$ and
$\mathbf{\Omega}=0.25,\,0.50$ and $1.0$, respectively. A smaller
Rossby number implies a larger rotation rate and the penetration
distance increases with decrease in the Rossby number from $0.407$
PSH for model R10 to $0.446$ PSH for model R13. For this set of
models with $\theta=0\,^\circ$, we see a relationship $\rm\Delta_p
\sim 0.38\, Ro^{-0.059}$.

Fig. 5(b) shows the variation of $\rm\Delta_p$ with $Ro$ for the
second set of models R3, R7 and R12 having $\theta=45\,^\circ$ and
$\rm\Omega\,=0.25,\,0.50$ and $1.0$, respectively. For this set, the
penetration distance decreases with the decrease in Rossby number.
Model R3 with  $Ro = 0.255$ has $\Delta_p = 0.399$ PSH, while R12
with $Ro = 0.069$ has $\Delta_p = 0.363$ PSH. This set of models
shows a scaling relation of $\rm\Delta_p \sim 0.44\, Ro^{0.072}$.

\subsection{Dependence of penetration height on colatitude $\theta$}

A set of five models R10, R2, R3, R4 and R5 has been examined to
study the effect of co-latitude ($\theta$) on the penetration
distance. All the models have {\boldmath$\rm\Omega$}$\,=0.25$ while
$\theta$ is changed from $0\,^\circ$ (Model R10) to $90\,^\circ$
(Model R5) in steps of $22.5\,^\circ$. Figure 4(b) shows the plot of
$\rm\Delta_p$ against $\theta$ for these five models. At
$\theta=0\,^\circ$, the penetration distance is maximum $0.407$ PSH.
It decreases slightly for successive values of $\theta$, and for
$\theta=90\,^\circ$ (case R5) has the minimum value of $0.210$ PSH.

Singh et al. (1994) computed many non-rotating cases having
different input fluxes and found a scaling relationship between the
upward penetration and the input flux ($F_b$) of the form $\Delta_p
\sim\, (\rm{F_b}/\rho_{ctop})^{1/3}$, where $\rho_{ctop}$ is the
density at the top of the convection zone. We attempt to find a
similar relationship which includes the effects of rotation.

In Fig. 6(a), we have plotted
($\rm{F_b}/\rho_{ctop})^{1/3}/\Delta_p$ for
{\boldmath$\rm\Omega$}$\,=0.25$, $0.5$, and $1.0$ corresponding to
models R10, R11, and R13, respectively. We find a scaling
relationship $\Delta_p \sim\,
(\rm{F_b}/\rho_{ctop})^{1/3}/0.283(\Omega)^{-0.065}$ for the three
cases having $\theta=0\,^\circ $. For cases R3, R7, R12 having
$\theta=45\,^\circ$ (Fig. 6(b)), we have a relationship of the form
$\Delta_p \sim\, (\rm{F_b}/\rho_{ctop})^{1/3}/0.35(\Omega)^{0.068}$.

For a set of models with a particular value of $\theta$, we have

\begin{eqnarray}
\Delta_p = c_1Ro^{x_1} \,\,\,\,\,\ \rm{and} \,\,\,\,\,\, \Delta_p =
\frac{(F_b/\rho_{ctop})^{1/3}}{c_2\Omega^{x_2}},
\end{eqnarray}
where $c_1$ and $c_2$ are constants and $x_1$ and $x_2$ are the
scaling powers for $Ro$ and $1/\Omega$.

One can use relations (4) to compute back the power laws of Fig. 6
by writing

\begin{eqnarray}
\Omega^{x_2} = \frac{(F_b/\rho_{ctop})^{1/3}}{c_1c_2Ro^{x_1}},
\end{eqnarray}
\begin{eqnarray}
x_2 =
\frac{\ln\frac{(F_b/\rho_{ctop})^{1/3}}{c_1c_2Ro^{x_1}}}{\ln\Omega}.
\end{eqnarray}

Substituting the value of $Ro,\, c_1,\,c_2,\,x_1, \rm{and}\,\,
\Omega\,$ in Eqn.(6), we find that for cases with
$\theta=0\,^{\circ}$, $x_2$ to be $-0.063$ and for cases with
$\theta=45\,^{\circ}$ it is $0.062$ which are close to the values
$-0.065$ and $0.068$, respectively, as shown in Fig. 6.

\begin{figure}
\centering{\includegraphics[width=8cm]{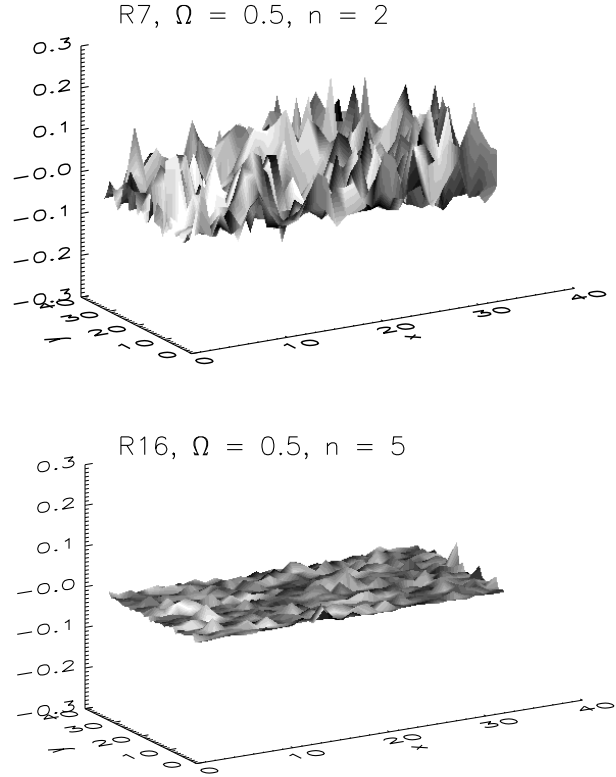}}
\caption{Instantaneous vertical velocity at a horizontal surface at
unstable-upper stable layer boundary located at a height of 0.6 from
the bottom for two cases R7 ($\mathbf{\Omega}=0.5,\,
\theta=45\,^\circ$) and R16 ($\mathbf{\Omega}=0.5,\,
\theta=45\,^\circ$). Polytropic indices of the upper stable layer
are 2 (Model R7) and 5 (Model R16). \label{fig.7}}
\end{figure}

\subsection{Dependence of penetration height on relative stability parameter $S$}

We have constructed four models with different stability parameter
($S$) defined in Eqn. (3) by varying the polytropic index of the
upper stable layer. We have $S = 1,\,3,\,5$ and $7$ for cases R7,
R14, R15 and R16, respectively. In Fig. 2(c), we have plotted the
distribution of the kinetic energy flux with height for these four
models. The values of the angular velocity ({\boldmath$\rm\Omega$})
and angle between the rotation vector and vertical axis ($\theta$)
are fixed at $0.5$ and $45\,^\circ$, respectively, for these models.
The corresponding penetration heights $\rm\Delta_u$ and
$\rm\Delta_p$ (in PSH) are given in the last two columns of Table 3.
Figure 2(c) clearly shows the differences in the kinetic flux
profile with $S$ especially in the region near the stable-unstable
layer interface. We find a decrease in penetration distance from
0.385 PSH in model R7 ($S=1$) to 0.058 PSH in model R16 ($S=7$).
Figure 7 shows the instantaneous vertical velocities at the
horizontal interface of the unstable-stable interface for models R7
and R16. The vertical velocities for model R16 with stiffer stable
layer (S=7) are much lower compared to that of model R7 (S=1).

Fig. 8 shows the variation of penetration distance with stability
factor $S$ for models R7, R14, R15, \& R16. These models show a
scaling relationship $\Delta_p \sim S^{-1}$ . This means that the
convection is efficient enough to establish a nearly adiabatic
stratification in the upper stable layer and the penetration is
nearly adiabatic (Hurlburt et al. 1994, Zahn 2002). Brummel et al.
(2002) and Pal et al. (2007) found a scaling relationship of
$\Delta_p \sim S^{-1/4}$ for downward penetration for rotating
convection zones associated with a thermal adjustment region at the
bottom.

\begin{figure}
\centering{\includegraphics[width=8cm]{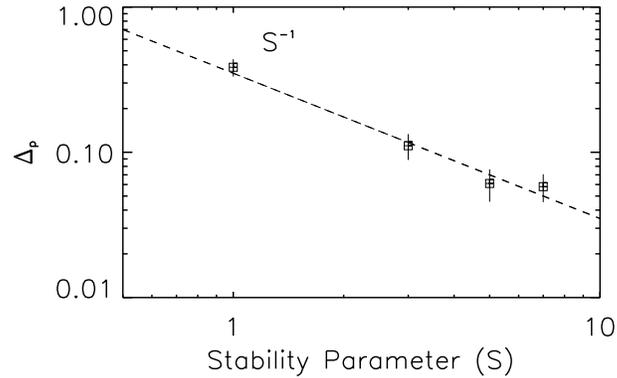}}
\caption{Variation of penetration distance $\rm\Delta_p$ with
stability parameter $S$ for models R7, R14, R15 and R16. Here,
$\mathbf{\Omega}=0.5$ and $\theta=45\,^{\circ}$. The dashed line
represents scaling law, $S^{-1}$. \label{fig.8}}
\end{figure}
\begin{table}
\caption{Parameters of models computed to examine the effect of
resolution on the upward penetration}\label{table:4} \centering
\begin{tabular}{ccccccc}
\hline\hline Models & Grid &{\boldmath$\rm\Omega$}  & $\theta/\pi$ &
$\theta$ &$t$ & $\Delta_p$ (PSH)\\
\hline
R1A & $64 \times 64 \times 128 $& 0    & 0   & $0^o$    &1056 & 0.303\\
R7A & $64 \times 64 \times 128$ &0.50 & 1/4 & $45^o$   & 907 & 0.333\\
R7B & $96 \times 96 \times 96$ &0.50 & 1/4 & $45^o$   & 550 & 0.268
\\ \hline
\end{tabular}
\end{table}

\subsection{Dependence of penetration height on resolution}

Table 4 lists additional models computed to see the effect of
resolution on the upward penetration. Model R1A is same as model R1
except that it has a total of $524,288\,(64 \times 64 \times 128)$
grid points instead of $117,600\,(35 \times 35 \times 96)$ grid
points, having better horizontal as well as vertical resolution.
Penetration height for non-rotating model R1 is $0.384$ PSH (cf.
Table 3), while for the non-rotating model R1A with increased
resolution, it is $0.303$ PSH.

Two rotating models R7 and R7A examine the effect of change in both
horizontal and vertical resolution on the penetration distance. We
notice a decrease in penetration depth from $0.385$ PSH in model R7
($35 \times 35 \times 96$) to $0.333$ PSH in model R7A ($64 \times
64 \times 128$). It may be noticed that while models R1 and R7 had
almost same penetration distance, the decrease in $\Delta_p$ with
increased resolution is smaller for model R7A which has a higher
rotation rate compared to model R1A which is non-rotating. It seems
that the effect of increase in resolution on penetration is
different for models with different rotation rates.

Three models R7, R17, and R7B examine the effect of horizontal
resolution on the penetration distance above the convection zone.
While all the models have 96 grid points in the vertical, they
differ in their horizontal resolution. Model R7 has $35 \times 35$
grid points in the horizontal direction, model R17 has $46 \times
46$ and model R7B has $96 \times 96$ grid points. The penetration
distance decreases with the increase in horizontal resolution from
$0.385$ PSH (Model R7) to $0.340$ PSH (Model R17) to $0.268$ PSH
(Model R7B).

\section{Conclusions}

We have computed $17$ models and presented the results of our
three-dimensional numerical simulations of turbulent compressible
convection penetrating into a radiative envelope under the influence
of rotation. We find that the penetration distances for these models
lie in the range $0.058\,\rm{PSH} \le \rm\Delta_p \le 0.446\,
\rm{PSH}$.

Recently, Woo \& Demarque (2001) put an empirical constraint on the
convective core overshoot for intermediate-to-low mass stars by
using Roxburgh's integral constraint. They found that the proper
limit of core overshoot for these stars would be $15\%$ of the core
radius. In Table 5, we have shown this calculation for our 17
models. We have computed the penetration distance in PSH as a
percentage of the total size of the convection zone. The size of the
convection zone is taken from the last column of Table 1 while the
penetration height ($\Delta_p$) has been taken from the last column
of Table 3. As can be seen from Table 5, We find an upper limit on
the penetration into the upper radiative layer to be around
$18.5\%$.

\begin{table}
\caption{Penetration as a percentage of the size of the Convection
Zone }\label{table:5} \centering
\begin{tabular}{cccccccccccccc}
\hline\hline Model& R1 & R2 & R3 & R4 & R5 & R6 & R7 & R8 & R9 & R10
& R11 & R12 &R13  \\
 $\%\Delta_p$ & 16.0
& 16.6 & 16.6 & 14.8 & 8.8 & 17.6 & 16.0 & 15.0 & 6.0 & 17.0 & 18.5
& 15.1 & 18.6 \\
Model & R14 & R15 & R16 & R17 & & & & & & & & &  \\
$\%\Delta_p$ & 4.0 & 2.0 & 1.7 & 14.2 & & & & & & & & &   \\
\hline
\end{tabular}
\end{table}

For rotation about a vertical axis ($\theta=0\,^\circ$), the
penetration into the upper stable region increases as the angular
rotational velocity ({\boldmath$\rm\Omega$}) increases or the Rossby
number ($Ro$) decreases.  However, this trend is reversed for
rotation around an inclined axis. When the angle
($\theta=45\,^\circ$), the penetration distance into the radiative
envelope  decreases with increasing {\boldmath$\rm\Omega$} owing to
horizontal mixing. To see the effect of change of angle of
inclination on penetration, angle $\theta$ is systematically varied
from $0\,^\circ$ to $90\,^\circ$ in steps of $22.5\,^\circ$ for five
models with a fixed angular velocity {\boldmath$\rm\Omega$}$=0.25$.
We again find that the penetration distance decreases as the
colatitude $\theta$ is increased. We do not see this behaviour
changing even when the resolution is increased.

We also find that the penetration distance above the convection zone
obeys a scaling relation of the form $\Delta_p \sim S^{-1}$
appropriate for nearly adiabatic penetration even in the presence of
rotation. The present simulations need to  be extended to include
more realistic input physics and a higher resolution to enable us to
get a better insight into the dynamics of rotating convection near
the convective-radiative interface. Since the motivation of the
study was to look into the effect of varying rotation rates on the
penetration distance, we needed to compute several models. Due to
the restrictions in speed and memory, we were unable to increase the
resolution of all the models. We hope to achieve this in future
studies.

\begin{acknowledgements}
      PSP is grateful to University Grants Commission, India for a
      Senior Research Fellowship. HPS was supported by grants from
      Indian Space Research Organization and Deutscher Akademisher Austauschdienst
      (DAAD), Germany. KLC was partly supported by
      grants from Hong Kong Research Grant Council. Authors acknowledge
the use of High Performance Computing facilities at IUCAA, Pune,
India.
\end{acknowledgements}



\end{document}